\documentclass[twocolumn]{aastex631}
\usepackage{showyourwork}
\usepackage{amsmath}
\usepackage{bm}
\usepackage{natbib}

\newcommand\ghurl[0]{\url{https://github.com/tedjohnson12/bin-disk-paper}}
\newcommand\codeurl[0]{https://github.com/tedjohnson12/polar-disk-freq}

\definecolor{github-icon-clone}{rgb}{0.12,0.47,0.71}

\newcommand{\RR}[1]{{#1}} 
\newcommand{\RP}[1]{\RR{#1}} 

\DeclareMathOperator{\arctantwo}{arctan2}
\DeclareMathOperator{\arccsc}{arccsc}

\begin{document}

\title{The fraction of polar aligned circumbinary disks}

\author[0000-0002-1570-2203]{Ted M. Johnson}
\author[0000-0003-2401-7168]{Rebecca G. Martin}
\author[0000-0003-2270-1310]{Stephen Lepp}
\affiliation{Nevada Center for Astrophysics, University of Nevada, Las Vegas, 4505 South Maryland Parkway, Las Vegas, NV 89154, USA}
\affiliation{Department of Physics and Astronomy, University of Nevada, Las Vegas, 4505 South Maryland Parkway, Las Vegas, NV 89154, USA}
\author[0000-0002-4636-7348]{Stephen H. Lubow}
\affiliation{Space Telescope Science Institute, 3700 San Martin Drive, Baltimore, MD 21218, USA}

\begin{abstract}

    Circumbinary gas disks that are misaligned to the binary orbital plane evolve toward either a coplanar or a polar-aligned configuration with respect to the binary host. The preferred alignment depends on the dynamics of the disk: whether it undergoes librating or circulating nodal precession, with librating disks evolving to polar inclinations and circulating disks evolving to coplanar.
    We quantify the fraction of binary star systems whose disks are expected to have polar orbits $f_\text{polar}$, extending previous work to include disks with non-zero mass.
    Our results suggest that, for low mass disks, the polar fraction is highly sensitive to the distribution of binary eccentricity with a higher fraction expected for higher binary eccentricities, $f_{\rm polar}\sim e_{\rm b}$. However, for massive discs, the fraction is independent of the binary eccentricity and $f_{\rm polar}\approx 0.37$. The value of $f_\text{polar}$ is always reduced in a population with a greater preference for low initial mutual inclination.
    We also explore the consequences of the finite lifetime and non-zero radial extent of a real disk, \RR{both of which affect a disk's ability to complete its evolution to a stationary configuration.} Our findings can be used to make predictions given populations with well-understood distributions of binary eccentricity, initial mutual inclination, and disk angular momentum.
\end{abstract}


\section{Introduction}
\label{sec:intro}

Observed circumbinary planets -- those orbiting two stars -- have mostly been found to orbit in the same plane as their hosts \citep{doyle2011,orosz2012,welsh2012}. While this preference is the result of target selection effects (\citealt{martin2014}, see also \citealt{martin2015}, \citealt{martin2017a}), it is also representative of the fact that, for the disk progenitors of these planets \citep[from which the planets inherit orbital parameters, see][]{childs2021}, a coplanar alignment is one of two possible stable orientations \citep{bate2000,lubow2000}. The other configuration -- a polar disk -- is characterized by being inclined $90^\circ$ relative to the binary orbital plane \citep{aly2015,martin2017}. \RR{O}nly a handful of confirmed polar disks have ever been discovered. These include a polar debris disk in the 99 Her system (\citealt{kennedy2012}, which likely evolved from a polar gas disk, see \citealt{smallwood2020}) and polar gas disks in the HD 98800 \citep{kennedy2019} and V773 Tau \citep{kenworthy2022} systems.

\RR{Recently, strong evidence for the first polar planet has been presented by \citet{baycroft2025}. They use radial velocity measurements of an eclipsing binary to infer retrograde apsidal precession which can be best explained by a planetary mass companion in a polar orbit.
The only previous evidence for a polar circumbinary planet comes from AC Her, a post-AGB star system that hosts an unusually-truncated circumbinary disk, with a planet located interior to the disk's inner edge being the most likely cause \citep{hillen2015,anugu2023,martin2023}.
}

Through surveys of recent literature in circumbinary disk characterization \citep{czekala2019,zurlo2023}, \citet{ceppi2024} estimate that the fraction of circumbinary disks which are polar ($f_\text{polar}$) is $0.17 \pm 0.08$, though they acknowledge that the sample size is small ($\sim 15$ disks) and that the mutual inclination is hard to measure. They place this result in the context of a theoretical framework for a massless disk \citep{farago2010,lubow2018,zanazzi2018} to conclude that observations are not telling the whole story. In this work, we use the results of \citet{martin2019} to extend this theoretical framework to massive disks.

\RR{Disks which are slightly misaligned from coplanar (or retrograde coplanar)  undergo circulating nodal precession about the binary angular momentum vector \citep{papaloizou1995}. \citet{bate2000} and \citet{lubow2000} show that this kind of motion leads to coplanar alignment {(or retrograde coplanar alignment)} on a timescale that scales as $h^2/\alpha$, where $h$ and $\alpha$ are the disk aspect ratio and viscosity parameter, respectively. Note that we use the terms prograde and retrograde to describe the direction of the orbit of the third body relative to the binary orbit, rather than the direction of the nodal precession. }

When the mutual inclination is $\sim\pi/2$ (near polar), the disk can librate about the eccentricity vector of the binary \citep[e.g.,][]{verrier2009,farago2010,doolin2011}. In \RR{addition, for} sufficiently massive disks, a third kind of motion arises at high inclinations. 
\RR{{\it Crescent} orbits take over at inclinations between libration and retrograde circulation, named because of the shape they form on a phase diagram  (\citealt{chen2019}, see also \citealt{zanazzi2018}). }
\citet{abod2022} showed that both librating and crescent orbiting disks decay to the stationary polar inclination \citep[see][]{martin2019} on an evolutionary timescale similar to that of coplanar alignment.

Thus, when the disk lifetime is longer than the alignment timescale, it is appropriate to map the secular dynamics (which we call the dynamical state, e.g. nodal precession, libration or crescent orbits) to an end state of the disk (coplanar or polar, respectively). More challenging is mapping the parameters of the disk-binary system (e.g., binary eccentricity, initial inclination, disk mass, etc.) to a dynamical state. In general, this can be done using expensive 3D disk simulations (e.g., those shown in \RR{\citealt{aly2015}, \citealt{cuello2019}} \citealt{martin2019}, \citealt{smallwood2019}, \citealt{abod2022}, \RR{\citealt{young2023} \& \citealt{ceppi2023}}). It is not feasible to perform such simulations over a large parameter space, so we instead approximate a disk as a geometrically narrow ring with angular momentum equivalent to that of the extended disk it represents. The literature provides a robust theoretical framework for the dynamics of such a ring, with analytical solutions \citep{farago2010,zanazzi2018,martin2019} as well as numerical $n$-body simulations \citep{abod2022,lepp2022} used to translate initial conditions into the ring's dynamical state. For simplicity, we assume that the narrow ring model approximates an extended disk sufficiently well. This is justified if warps and breaks are not important to the disk's evolution, though we discuss these effects in the context of our model.


In this study we use three methods, varying from purely analytic to $n$-body and statistical, to calculate $f_\text{polar}$ for a given population. In Section \ref{sec:analytic} we extend the analytic work to massive rings. Section \ref{sec:rk} 
describes the numerical integration of a set of coupled differential equations that describe the evolution of the binary-ring system in the secular approximation of the binary potential. In Section \ref{sec:reb} we treat the binary potential fully using the $n$-body code {\sc rebound} \citep{rebound}, finding only a negligible difference between this method and the much-faster secular approximation. We discuss our results in Section \ref{sec:results}, specifically looking at the effect that the binary eccentricity, relative angular momentum, and initial mutual inclination distributions have on the value of $f_\text{polar}$. Section \ref{sec:discussion} examines the consequences of our narrow-ring approximation, and discusses what considerations must be made to apply our results to an extended disk. Finally, Section \ref{sec:conclusions} describes our overall conclusions.

\RR{
Code \RR{that implements} the methods described in this paper is publicly available\footnote{\url{\codeurl}, or via \hfill \break \texttt{ pip install git+\codeurl} } as an installable Python package \RR{\citep[\texttt{v1.0.1}]{johnson2025a}}. The \LaTeX\space source code can be found at \dataset[doi: 10.5281/zenodo.15347131]{\doi{10.5281/zenodo.15347131}} and the scripts and data used to generate each figure can be found by clicking on the \textcolor{github-icon-clone}{\GitHubIcon} and \textcolor{github-icon-clone}{\faDatabase} symbols, respectively.}
\section{Analytic Treatment}
\label{sec:analytic}
For a massive ring that is misaligned to the binary orbit there are two conditions that determine if the ring is librating \citep[see Equations 31, 32, 38 in][]{martin2019}. Each condition is valid for a different regime of the constant of motion $\chi$ that is defined with
\begin{equation}
    \chi = e_{\rm b}^2 - 2\,(1 - e_{\rm b}^2)\,j\,(2j + \cos{i})\, ,
\end{equation}
for binary eccentricity $e_{\rm b}$, normalized ring angular momentum $j = J_{\rm r}/J_{\rm b}$, and mutual inclination $i$. The condition for libration is
\begin{equation}
    \label{eq:lib_condition}
        \begin{array}{lr}
            -(1-e_\text{b}^2)(2j+\cos{i})^2 + 5e_\text{b}^2 \sin^2{i}\sin^2{\Omega} > 0\, , &\text{if}~ \chi > 0\, , \\
            e_\text{b}^2 + 4j(1-e_\text{b}^2)(-\cos{i} + j(-2+5\sin^2{i}\sin^2{\Omega})) > 0\, , &\text{if}~ \chi < 0\,.
        \end{array}
\end{equation}
These conditions can be rearranged to find the minimum value of $\Omega$ that will allow for a librating orbit.
\begin{equation}
    \label{eq:omega_min}
    \sin^2{\Omega_{\rm min}} = 
    \left \{
    \begin{array}{lr}
         \frac{(1-e_{\rm b}^2) (2j + \cos^2{i})}{5 e_{\rm b}^2 \sin^2{i}} & \text{if} ~\chi \ge 0\, , \\
    \frac{2}{5 \sin^2{i}} + \frac{\cos{i}}{5j \sin^2{i}} - \frac{e_{\rm b}^2}{20 j^2 (1-e_{\rm b}^2) \sin^2{i}} & \text{if} ~\chi < 0\, .
    \end{array}
    \right .
\end{equation}

Since the range of $\sin^2{\Omega_{\rm min}}$ is $[0,1]$, there exists a critical value for the inclination, $i_{\rm crit}$, below which all orbits circulate. For the $\chi \ge 0$ case, we require
\begin{equation}
    \label{eq:icrit1}
    \frac{5 e_{\rm b}^2}{1-e_{\rm b}^2} \sin^2{i_{\rm crit}} - (\cos{i_{\rm crit}} + 2j)^2 = 0\, ,
\end{equation}
and for the $\chi < 0$ case, $i_{\rm crit}$ satisfies
\begin{equation}
    \label{eq:icrit2}
    \sin^2{i_{\rm crit}} - \frac{\cos{i_{\rm crit}}}{5j} + \frac{e_{\rm b}^2}{20 j^2 (1-e_{\rm b}^2)} + \frac{2}{5} = 0\, .
\end{equation}
If the left-hand side of equation~(\ref{eq:icrit1}) or~(\ref{eq:icrit2}) (depending on the value of $\chi$) evaluates to $<0$, then only circulating orbits are possible. Note that when $j=0$, $\chi \ge 0$ and equation (\ref{eq:omega_min}) recovers equation (75) from \citet{zanazzi2018}.

Given a uniformly distributed $\Omega$, the probability of a polar orbit  is given by
\begin{equation}
    \label{eq:prob_polar}
    P(j,e_{\rm b}, i) = 
    \left \{
    \begin{array}{lr}
         1 - \frac{2}{\pi} \Omega_{\rm min} & \text{if} ~i > i_{\rm crit} \\
        0 & \text{if} ~i \le i_{\rm crit}
    \end{array}
    \right .
\end{equation}
\citep{zanazzi2018,ceppi2024}.
The fraction of polar orbits is then calculated with
\begin{equation}
    \label{eq:fp_int}
    f_{\rm polar} = \iiint P(j,e_{\rm b}, i)\,p_j(j)\,p_{e_\text{b}}(e_\text{b})\,p_i(i)\, dj\,de_{\rm b}\,di\,,
\end{equation}
where $p_j(j)$, $p_{e_\text{b}}(e_\text{b})$, and $p_i(i)$ are the probability distributions of $j$, $e_\text{b}$, and $i$, respectively.

This integral is straightforward to compute numerically, and allows for any desired distribution in $j$, $e_{\rm b}$, or $i$, so long as they are normalized. For example, an isotropic distribution of disk orientation, relative to the binary, would require $p_i(i) = \frac{1}{2}\sin{i}$ for $i$ in the range of $0$ to $\pi$.

\subsection{High-$j$ limit}
\label{subsec:high_j}

When $j$ is very large, the $\chi < 0$ branch of Equation \ref{eq:omega_min} becomes
\begin{equation}
    \sin^2{\Omega_{\rm min}} = \frac{2}{5}\csc^2{i}\, ,
\end{equation}
and the value of $i_{\rm crit}$ is 
\begin{equation}
    i_{\rm crit} = \arccsc{\sqrt{5/2}} = 39.2^\circ\,.
\end{equation}
This is the inclination of the last circular orbit for Kozai-Lidov oscillations in the limit that the perturbed object is very far from the perturber \citep{vonzeipel1910,kozai1962,lidov1962}. Note that it has no dependence on the disk angular momentum or on the binary eccentricity. 

For an isotropic \RR{distribution of the angular momentum vector direction (i.e. one where the specific angular momentum $\bm{\ell}$ has no preference for any direction)}, the polar fraction in this limit is
\begin{align}
    f_{{\rm polar}, j \gg 1} &= \int_0^\pi\,\left (1-\frac{2}{\pi}\arcsin{\left (\sqrt{\frac{2}{5}} \csc{i} \right )}\right ) \sin{i}\,di\, , \\
    & = %
  0.37\unskip\label{output/high_j_integral.txt}\unskip%
 \, . \nonumber
\end{align}
\RR{This is the value that our calculations in Section \ref{sec:rk} tend to for high $j$ and low $e_\text{b}$}

To interpret the Kozai-Lidov oscillations in this limit we can picture a system from the point of view of a far-away stationary observer. When $j \approx 0$, the orbital parameters of the interior binary are not affected by the disk, and the disk undergoes precession or libration due to the effects of the binary. When $j$ is high, however, it is the disk that appears static. An observer would see oscillations in the orbital parameters of the central binary, and its angular momentum and eccentricity vectors would change in response to the disk. This can be thought of in terms of the classical Kozai-Lidov problem, where instead of a planet and its satellite, the interior objects are the binary, and the exterior perturbing object is the circumbinary disk.

\section{Secular Approximation}
\label{sec:rk}
\citet[Equations (7-10)]{martin2019} also provide a set of coupled first-order differential equations that describe the evolution of $e_{\rm b}$ along with the direction of the disk angular momentum $(\ell_x, \ell_y, \ell_z)$, based on previous work by \citet{farago2010}. We solve these equations using the Runge-Kutta-Fehlberg (RKF, \citealt{fehlberg1969}, see \citealt{hairer2000} p. 177 for implementation) -- a numerical method similar to the common 4th order Runge-Kutta integrator, but that computes a 5th order solution to allow for a variable step size. This method treats the binary in the quadrupole approximation, meaning that it is not sensitive to the effects of the binary mass fraction or orbital dynamics that occur on timescales less than the binary orbital period \citep[e.g.][]{naoz2016}. However, the computational cost is extremely low; we are able to run $10^4$-$10^5$ simulations per second -- about 10,000 times faster than running $n$-body simulations (see Section \ref{sec:reb}).

We determine the dynamical state of the system by tracking the orbital quantities $x=i\,\cos{\Omega}$ and $y=i\,\sin{\Omega}$, where
\begin{equation}
    i = \arccos{\ell_z}
    \label{eq:inclination}
\end{equation}
and
\begin{equation}
    \Omega = \arctantwo (\ell_x, -\ell_y)\, .
    \label{eq:omega}
\end{equation}
\RR{If a disk undergoes circulating nodal precession, then the path $(x,\,y)$ forms closed loops about the origin as the system evolves. }
However, any state that would lead to a polar orbit (i.e. libration and crescent patterns) will form closed loops that do not encompass the origin, and these tracks will never cross the line $i\,\sin{\Omega}=0$. The state can be determined by tracking the quadrant \RR{(defined using the standard counter-clockwise convention)} of the point $(i\,\cos{\Omega},~i\,\sin{\Omega})$. \RP{Circulating orbits will traverse all four quadrants.}
A polar orbit that starts in quadrant I or II, however, is restricted to those two quadrants and will alternate between them. The same goes for polar orbits in quadrants III and IV. Therefore, the dynamical state can often be determined by the list of previous quadrant crossings.

\begin{figure}
    \begin{centering}
        \includegraphics[width=0.5\textwidth]{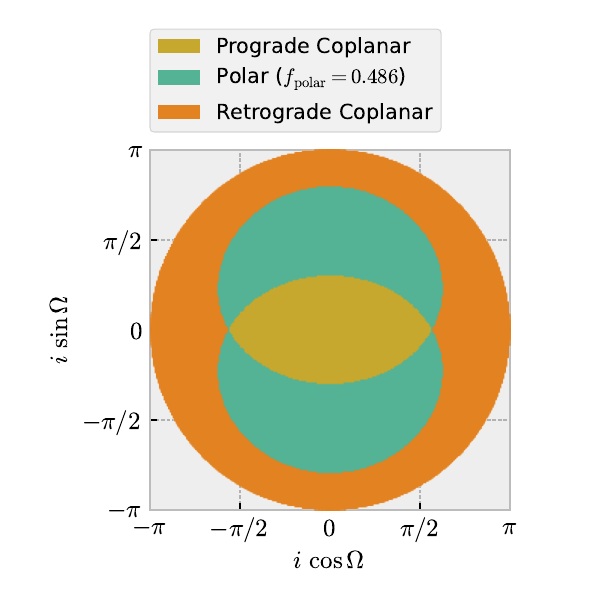}
    \end{centering}
    \caption{
        Grid of dynamical states for %
  \input{output/grid_example_n_sims.txt}\unskip\label{output/grid_example_n_sims.txt}\unskip%
 RKF simulations of a system with $e_{\rm b}=%
  \input{output/grid_example_eb.txt}\unskip\label{output/grid_example_eb.txt}\unskip%
$ and $j=0.1$. This grid can be run for any combination of $(e_{\rm b},\,j)$, and integrating over the polar region computes the fraction of systems in which a disk would evolve to a polar state.
    }
    \script{grid_example.py}
    \label{fig:grid_example}
\end{figure}

We estimate the value of Equation (\ref{eq:fp_int}) using this method by computing the integral of $P(j,e_{\rm b},i)\,p_i(i)\,di$ over all $i$; this is the probability of a polar orbit given some $j$ and $e_{\rm b}$. This is done numerically using a grid of simulations, as shown if Figure \ref{fig:grid_example}. These simulations all vary in their initial conditions, with the longitude of the ascending node $\Omega \in [0,2\pi)$ and mutual inclination $i \in [0,\pi]$.

\section{$n$-body simulations}
\label{sec:reb}

We use the N-body integration software {\sc rebound} \citep{rebound} to simulate a circumbinary system with an additional level of sophistication compared to those described in Section \ref{sec:rk}, in which the binary was treated in the quadrupole approximation. We model the disk as a point mass with its same orbital parameters and angular momentum.

\begin{figure}
    \begin{centering}
        \includegraphics[width=0.5\textwidth]{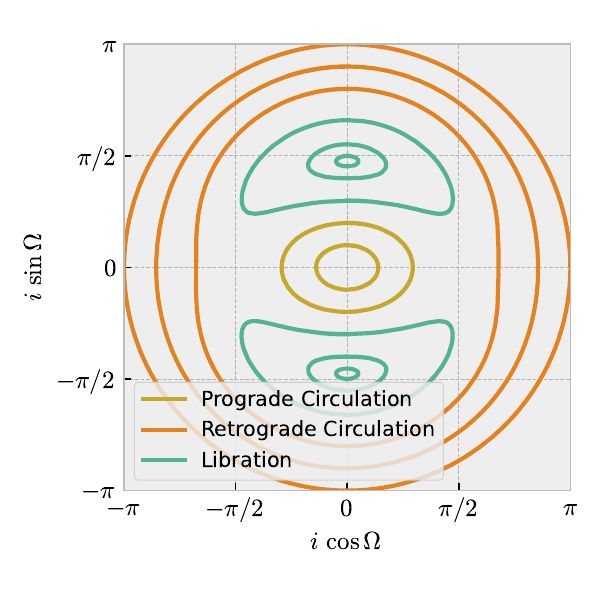}
        \caption{Example of system simulation and state determination. This figure tracks the orbital parameters
        of a third body with $j=0.05$ orbiting a $1 M_\odot$ binary with equal mass stars and $e_b = 0.4$.
        Initially $\Omega = \frac{\pi}{2}$ in all cases.}
        \label{fig:sim_example}
        \script{sim_example.py}
    \end{centering}
\end{figure}

For any given system, the dynamical state is determined similarly to the method described in Section \ref{sec:rk}. Note, however, that the $(i\,\cos{\Omega},~i\,\sin{\Omega})$ tracks now contain small structure on the orbital timescale of the outer mass. These structures can mean that a simulation may not make monotonic progress along its track, occasionally backtracking. Such a backtrack can give a false reading of the dynamical state if it occurs across the $x$ or $y$ axis. To mitigate this problem, we integrate each system until it returns to its original position in $(i\,\cos{\Omega},~i\,\sin{\Omega})$ space. If, at this point, the path does not enclose the origin, then it is polar. \RR{Prograde and retrograde circulating orbits can be distinguished by the initial inclination.} Figure \ref{fig:sim_example} shows paths for a set of simulations initialized with $\Omega = \pi/2$.

We use the IAS15 Gauss-Radau integrator \citep{reboundias15}, which is a 15th-order variable step size numerical integrator. The maximum time-step this requires is on the order of the binary orbital period, and the integration is therefore much slower than the RKF algorithm, at approximately 2 iterations per second (for $r/a_\text{b}=5$, $r$ is the radius of the ring). However, the time required to determine the dynamical state depends on the precession timescale, with some simulation states still undetermined after $10\,000$ outer-mass orbits ($\sim \,$5 seconds). This happens rarely enough that it is most efficient to simply stop the integration, and return an unknown state which must be factored in to our uncertainties. The precession timescale is greatest near the boundary of the precession and libration regions (formally infinite at the boundary, \citealt{farago2010}, see also Figure 16 of \citealt{rabago2024}), and so these boundary simulations are the most likely to be stopped before a dynamical state can be determined. We therefore expect both polar and coplanar simulations to be terminated early at approximately equal rates, and that this method will not introduce bias into our results.

Given that these simulations take much longer than the RKF algorithm, we cannot compute the same grid of dynamical states. Instead, we employ a Monte Carlo (MC) method to estimate the probability of a polar orbit. For any set of parameters (including the binary fraction $f_{\rm b}$ now) we sample isotropically on the $i$-$\Omega$ sphere, and use the bootstrap method to estimate $f_{\rm polar}$. We sample in batches of 4 and stop when the bootstrap confidence interval metric reaches a prescribed value. After each batch, the results are written to an SQLite database to be recalled in the future. Upon initialization, the MC sampler queries this database for all relevant results.

\begin{figure}
    \begin{centering}
        \includegraphics[width=0.5\textwidth]{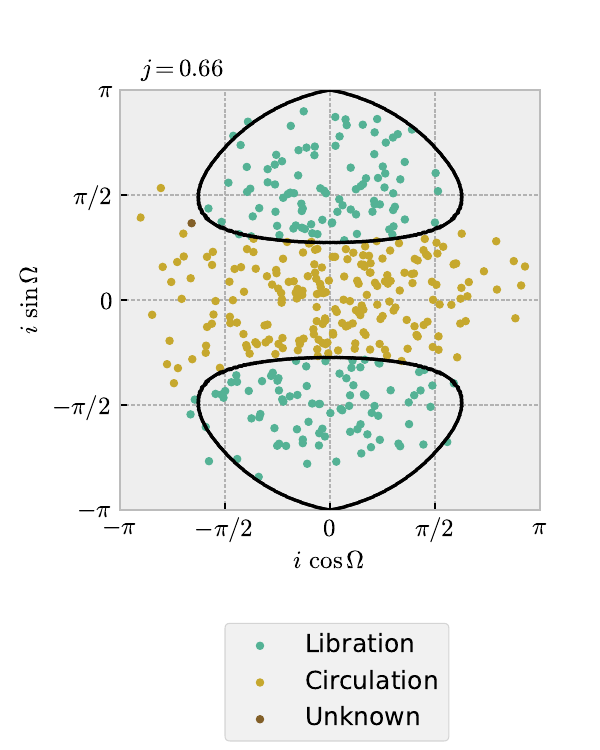}
        \caption{
            MC results (points) for $e_{\rm b}=0.4$, $j=0.66$, and $f_{\rm b}=0.5$ compared to RKF results for the same setup (contour lines).
            \RR{At this value of $j$ all circulating orbits have the same secular behavior regardless of the orbit being prograde or retrograde.}
             The regions interior to the black contour lines are initial conditions which allow polar \RR{librating} orbits according to the RKF method.
            Note that some points that {\sc rebound} classifies as polar fall outside the RKF polar region.
            The MC result gives $f_{\text{polar}} = %
  \input{output/mc_out.txt}\unskip\label{output/mc_out.txt}\unskip%
$ while the RKF result gives $f_{\text{polar}} = %
  \input{output/rk_out.txt}\unskip\label{output/rk_out.txt}\unskip%
$ with
            numerical error on the order of $%
  \input{output/rk_err.txt}\unskip\label{output/rk_err.txt}\unskip%
$. \RR{The point marked ``Unknown'' represents a system whose state could not be determined after $10\,000$ orbits of the third body due to a relatively long dynamical timescale and has been factored into the MC uncertainty.} \RR{The MC simulation results used to generate this figure can be downloaded at \dataset[doi: 10.5281/zenodo.14736363]{\doi{10.5281/zenodo.14736363}}}
        }
        \script{compare_massive.py}
        \label{fig:rkf}
    \end{centering}
\end{figure}

Figure \ref{fig:rkf} shows that, while the RKF and {\sc rebound} simulations produce similar results, they are not identical. Specifically noticeable are points sampled by the MC algorithm that lie outside the RKF libration region (with a higher inclination), but are found by {\sc rebound} to librate. There are also points with the opposite discrepancy -- those that lie in the RKF libration region but produce \RP{circulation} -- but they are fewer and not as noticeable by eye. This discrepancy arises because the secular approximation averages the central potential over the orbital period of the binary, while in this section we are treating the full $n$-body dynamics. The two schemes converge when $r \gg a_\text{b}$, and we can eliminate this discrepancy by increasing $r$ (e.g., $r=25a_\text{b}$ compared to our nominal choice of $5a_\text{b}$). In general, the choice of small $r/a_\text{b}$ changes the shape of the parameter space that allows polar orbits when compared to those found in Section \ref{sec:rk}, but not enough to have a significant effect on $f_\text{polar}$. In fact, this bias is small enough that -- given uncertainties introduced by our limited number of samples -- we cannot distinguish between the polar fractions inferred from the two methods. Because the secular approximation used in Sections \ref{sec:analytic} \& \ref{sec:rk} require significantly lower computational cost, and the differences in results are insignificant, we choose to focus on the secular approximation results in the following sections.

\section{Results}
\label{sec:results}

To determine the polar fraction of a given population of geometrically narrow rings we must consider three distributions: $p_j(j)$, $p_{e_\text{b}}(e_\text{b})$, and $p_i(i)$. If these distributions were known, it would be easy to calculate $f_\text{polar}$ for a population. However, none are well constrained, so in this section we calculate $f_\text{polar}$ considering one distribution at a time, drawing on parameterizations from relevant literature when available.

\subsection{Effects of $j$}
\label{subsec:effect-j}

\begin{figure*}[!htbp]
    \begin{centering}
        \includegraphics[width=1\textwidth]{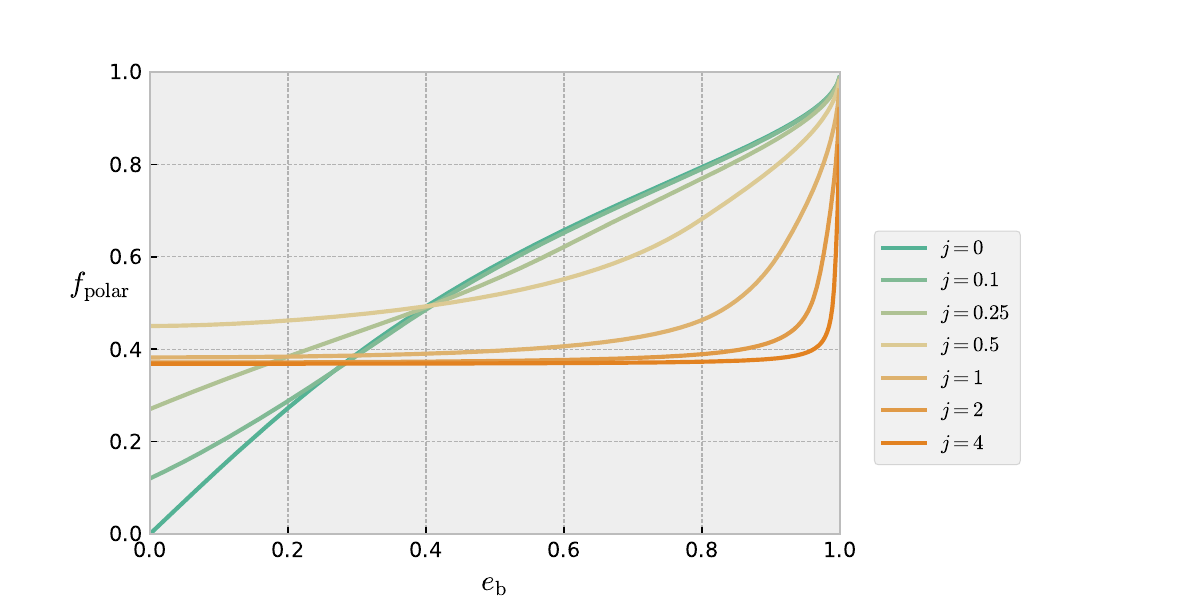}
        \caption{
            Polar fraction as a function of $e_{\rm b}$ for various values of $j$. Each line is computed via a numerical integration of equation (\ref{eq:prob_polar}), assuming an isotropic distribution of $i$.
       For low-$j$, the polar fraction is a strong function of $e_{\rm b}$. However, as $j$ increases, $f_{\rm polar}$ becomes insensitive to the binary eccentricity (except for the case that $e_\text{b}\approx 1$). Note that $f_\text{polar}(e_\text{b}\approx 0)$ is maximized when $j=0.5$.
        }
        \script{compare_ebin.py}
        \label{fig:ebin}
    \end{centering}
\end{figure*}

We choose to consider specific values of $j$ rather than a parameterized distribution (equivalent to setting $p_j(j) = \delta(j-j_0)$). This choice is motivated by the lack of published observational or theoretical constraints on the disk-to-binary angular momentum ratio. Even if the distribution were known, it would likely not be useful because of the possibility for extended disks to warp and break, leading to independent ring evolution (see Section \ref{sec:discussion}).

Figure \ref{fig:ebin} shows $f_\text{polar}(e_\text{b})$ for various values of $j$. We see here two limiting behaviors: when $j$ is small, then $f_\text{polar}\sim e_\text{b}$. However, when $j$ is large (and $e_\text{b}$ is not close to $1$), then the Kozai-Lidov behavior discussed in Section \ref{subsec:high_j} dominates and $f_\text{polar}\approx \text{const}$. Note that, while these calculations assumed $p_i(i)$ such that the ring direction is isotropic, we show in Section \ref{subsec:change_i} that the qualitative behaviors are independent of any preference (or lack of preference) for coplanar initial conditions.

\subsection{Effects of $e_\text{b}$}
\label{subsec:effect-eb}

\begin{figure*}
    \centering
    \includegraphics[width=\textwidth]{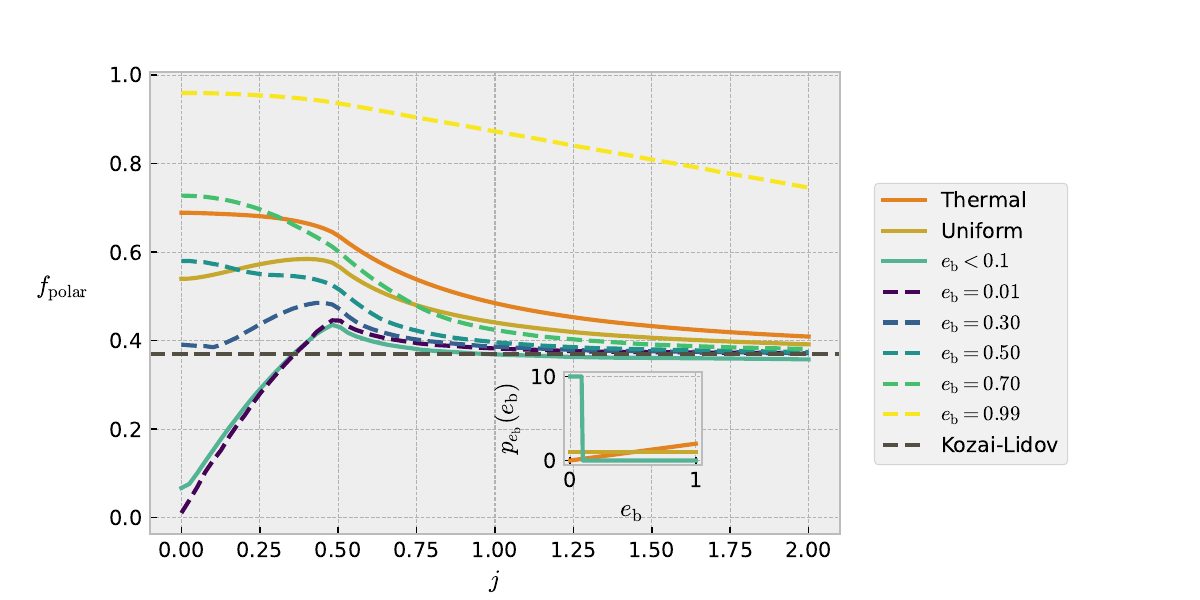}
    \caption{
        Polar disk fraction $f_\text{polar}$ as a function of $j$ for various distributions $p_{e_\text{b}}(e_\text{b})$. All cases assume isotropic distributions of initial angular momentum. In the thermal case, $p_{e_\text{b}}(e_\text{b}) \propto e_\text{b}$. In the uniform case, $p_{e_\text{b}}(e_\text{b}) = \text{const}$. The case labeled $e_\text{b} < 0.1$ uses a uniform distribution, but restricts the domain to $e_\text{b} \in [0,0.1)$. These three distributions are shown in the inset. The dashed line cases fix $e_\text{b}$ at a particular value. Notice the maximum that occurs at $j\approx 0.5$. This occurs because the appearance of crescent orbits at $j=j_\text{cr}$ \citep[eq. A3]{martin2019} make the parameter space that eventually evolves to polar very large.
    }
    \script{f_p_of_j.py}
    \label{fig:fpol_j}
\end{figure*}

Typical distribution functions for binary eccentricity have the form $p_{e_\text{b}}(e_\text{b}) \propto e_\text{b}^\alpha$, with the special cases of $\alpha=0,\,1$ referred to as ``uniform'' and ``thermal'', respectively \citep[see also \citealt{ceppi2024}]{hwang2022}. \citet{hwang2022} found that, for wide binaries, the distribution changes from uniform at $a_\text{b}\approx 100\text{AU}$ to thermal at $a_\text{b}\approx 500\text{AU}$. For shorter-period binaries, tidal forces lead to a strong preference for circular orbits, but it is not clear that the circularization timescale is short enough to use as an initial condition \citep[e.g.,][]{meibom2005,geller2012}. The requirement on the circularization timescale would be $t_\text{circ} < t_p$, where $t_p$ is the precession timescale.

Figure \ref{fig:fpol_j} shows $f_\text{polar}$ as a function of $j$ for various choices of $p_{e_\text{b}}(e_\text{b})$. As expected, the curves converge in the Kozai-Lidov regime, \RR{except for very large $e_{\rm b}=0.99$.} The uniform and thermal distributions both produce polar fractions $>0.5$ for small $j$, with the thermal distribution slightly higher. Except for when $j$ is close to 0, the circularized $e_\text{b} < 0.1$ curve follows the $e_\text{b} = 0.01$ curve very closely, including a maximum at $j=0.5$. This feature is present to some degree in all the low-$j$ curves, and is the result of the fact that, for $e_\text{b}=0$, the critical value for $j$ is 0.5 \citep{abod2022}. \RP{It is around this value of $j$ that crescent orbits begin to appear, and the size and shape of the libration region changes very rapidly. For general $e_\text{b}$, the transition occurs when $j\sim j_\text{cr}$, where $j_\text{cr}$ is the critical value of $j$ \citep[eq. A3]{martin2019}.}

\subsection{Effects of $i$}
\label{subsec:change_i}

\begin{figure*}
    \begin{centering}
        \includegraphics[width=\textwidth]{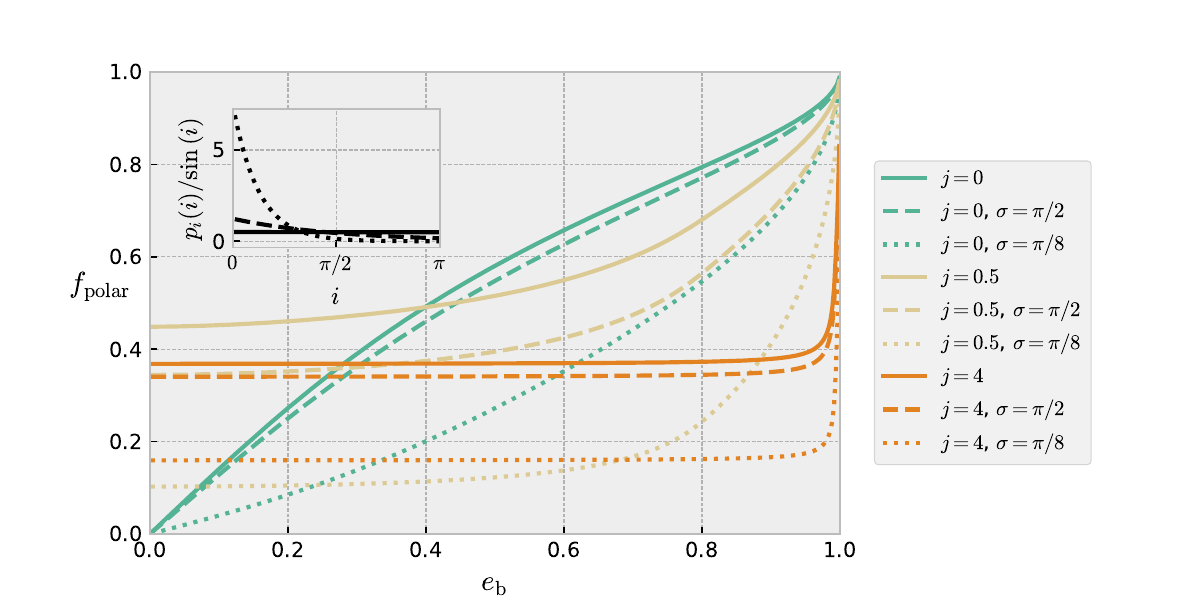}
        \caption{
            Same as Figure \ref{fig:ebin}, but with the mutual inclination distribution given by equation (\ref{eq:idist}) \RR{that is plotted in the inset normalized by $\sin i$. This normalization shows that the probability-per-unit-area on the unit sphere becomes very high at $i=0$ (coplanar alignment) when $\sigma$ is small (dashed ($\sigma=\pi/2$) and dotted ($\sigma=\pi/8$) lines), but it is uniform everywhere in the isotropic case ($\sigma \rightarrow \infty$, solid line). The main figure shows that for low $j$, low $\sigma$ (low inclination preference) decreases $f_\text{polar}$. For high-$j$, low $\sigma$ (low inclination preferences) lowers the value that $f_\text{polar}$ tends to as $e_\text{b}\rightarrow 0$.}
        }
        \script{compare_ebin_idist.py}
        \label{fig:ebin-idist}
    \end{centering}
\end{figure*}

\begin{figure*}
    \begin{centering}
        \includegraphics[width=\textwidth]{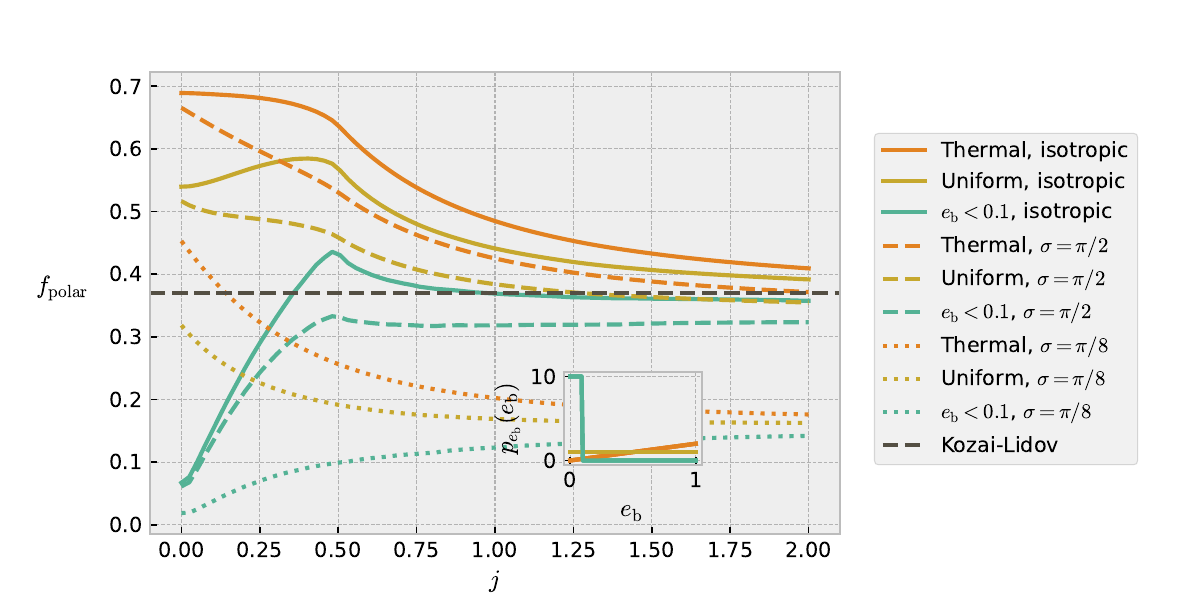}
        \caption{
            Same as Figure \ref{fig:fpol_j}, but with the mutual inclination distribution given by equation (\ref{eq:idist}). \RR{Preference for lower mutual inclinations (lower values of $\sigma$, see dashed and dotted lines) lead to lower $f_\text{polar}$ in all cases. See the inset of Figure \ref{fig:ebin-idist} for a visualization of the various $i$ distributions.}
        }
        \script{f_p_of_j_idist.py}
        \label{fig:fpol_j-idist}
    \end{centering}
\end{figure*}

In all previous results we have assumed that the there is no preferred direction of $\bm{j}$. For a population, this means that $\Omega$ is uniformly distributed between $0$ and $2\pi$ and $i$ is associated with the distribution function
\begin{equation}
    \label{eq:iso-i}
    p_i(i) = \frac{1}{2}\sin{i} .
\end{equation}

Hydrodynamic simulations by \citet{elsender2023} found that this assumption is valid for short-period binaries (those with separations $\sim 1$ AU). For wider binaries, however, a distribution function that prefers alignment would be more appropriate.

In this section we will break from the isotropic assumption and look at how $f_\text{polar}$ changes when we adopt a distribution that prefers low initial mutual inclinations. \citet{ceppi2024} proposed a normalized exponential distribution:
\begin{equation}
    p_i(i) = \frac{1}{\sigma(1-\exp{(-\pi/\sigma)})} \exp{\left(\frac{-i}{\sigma}\right)},
\end{equation}
where $\sigma$ is a parameter that encodes the distribution's preference (or lack of preference) for low mutual inclinations.
However, this distribution does not become isotropic for large $\sigma$, and results in probability densities (per unit solid angle) that diverge at $i=0,\,\pi$. We employ the following parameterization instead:
\begin{equation}
    \label{eq:idist}
    p_i(i) = \frac{1+\sigma^2}{\sigma^2 (\exp{(-\pi/\sigma)}+1)}\,\sin{i}\,\exp{\left(\frac{-i}{\sigma}\right)},
\end{equation}
which converges to equation (\ref{eq:iso-i}) in the limit $\sigma \rightarrow \infty$.

Figure \ref{fig:ebin-idist} shows that when low mutual inclinations are preferred as an initial condition, the fraction of systems that result in a polar orbit drop\RR{s} dramatically, and that this effect is fairly consistent across values for $e_\text{b}$. Figure \ref{fig:fpol_j-idist} similarly shows this effect, but with an interesting caveat: in the model with the heaviest preference for coplanar initial conditions, the fraction of polar disks is fairly high when $j=0$ and $e_\text{b}$ has either a uniform of thermal distribution ($>0.5$ in the thermal case).

\section{Discussion}
\label{sec:discussion}

\subsection{Timescales}
In order to map a dynamical state to a final alignment (e.g. librating orbit $\mapsto$ polar alignment), we are required to assume that the disk lifetime $t_\text{disk}$ is longer than the disk evolution timescale $t_\text{evol}$. The evolution timescale is the characteristic decay time for a disk whose inclination has been perturbed from a stable point (e.g., polar or coplanar), and while it is described by \citet{lubow2018} in their equation (29), it is only known up to a constant of proportionality:
\begin{equation}
    \label{eq:tevol}
    t_\text{evol} \propto \frac{h^2}{\alpha f^2(1-f)^2 e_\text{b}^2(1+4e_\text{b}^2) \Omega_\text{b}} ,
\end{equation}
where $h\equiv H/r$ is the disk aspect ratio, \RR{$f\equiv M_2/M_\text{b}$ is the mass ratio between the secondary star and the total binary mass}, and $\Omega_\text{b}$ is the binary orbital frequency. 

We can further constrain the evolution timescale if we decide that it must not be shorter than the precession timescale, and is likely of similar order. \citet[see eq. (2.32)]{farago2010} give an analytic expression for the precession period of a test particle orbiting an eccentric binary, which we write in our own variables as
\begin{equation}
    t_\text{p} = \frac{16}{3f(1-f)} \Omega_\text{b}^{-1} \left(\frac{r}{a_\text{b}}\right)^{7/2} \Gamma_1(e_\text{b},i,\Omega)\, ,
\end{equation}
where $\Omega_\text{b}$ is the binary orbital frequency and $\Gamma_1$ is a dimensionless constant that depends on the initial conditions:
\begin{equation}
    \Gamma_1(e_\text{b},i,\Omega) = \frac{K(k^2)}{\sqrt{(1-e_\text{b}^2)(h+4e_\text{b}^2)}} \, .
\end{equation}
$k^2$ and $h$ are constants that \citet{farago2010} define in their equations (2.31) and (2.20), respectively. $K(k^2)$ is the elliptic integral of the first kind, which \citet{farago2010} gives in equation (2.33).

Note that $\Gamma_1$ depends on $e_\text{b}$, and so it is only a constant in the case of a test particle. For $j>0$, we must compute the precession timescale numerically. The time coordinate used in Section \ref{sec:rk} is $\tau \equiv \alpha' t$. $\alpha'$ is defined by \citet[see eq. 3.9]{farago2010}, and we write it as
\begin{equation}
    \alpha' \equiv \frac{3}{4} \Omega_\text{b} \left(\frac{a_\text{b}}{r}\right)^{7/2} f(1-f) \,\Gamma_2(j,f,a_\text{b}/r, e_\text{b}) \, ,
\end{equation}
where we have written the $\sqrt{1+m_\text{r}/M_\text{b}}$ factor as $\Gamma_2$. \RR{$m_\text{r}$ is the mass of the ring. To compute $\Gamma_2$ from our variables requires finding the roots of a cubic equation (see equation (\ref{eq:gamma2-eq})). However, for low $j$ (e.g. $j<10$), $m_\text{r} \ll M_\text{b}$ and $\Gamma_2 \sim 1$.}

Given \RR{$\tau_\text{p}$, the precession timescale in units of $\tau$, we can calculate the precession timescale in physical units}
\begin{equation}
    \label{eq:tp-j}
    t_\text{p} = \frac{4}{3}\Omega_\text{b}^{-1} \left(\frac{r}{a_\text{b}}\right)^{7/2} \frac{\tau_\text{p}}{f(1-f)\Gamma_2}\, .
\end{equation}
Compared to the precession timescale for a test particle, that is
\begin{equation}
    \label{eq:tp-rat}
    \frac{t_{\text{p,}j}}{t_{\text{p,}j=0}} = \frac{\tau_\text{p}}{4\,\Gamma_1\,\Gamma_2} \, .
\end{equation}

\begin{figure}
    \includegraphics[width=0.5\textwidth]{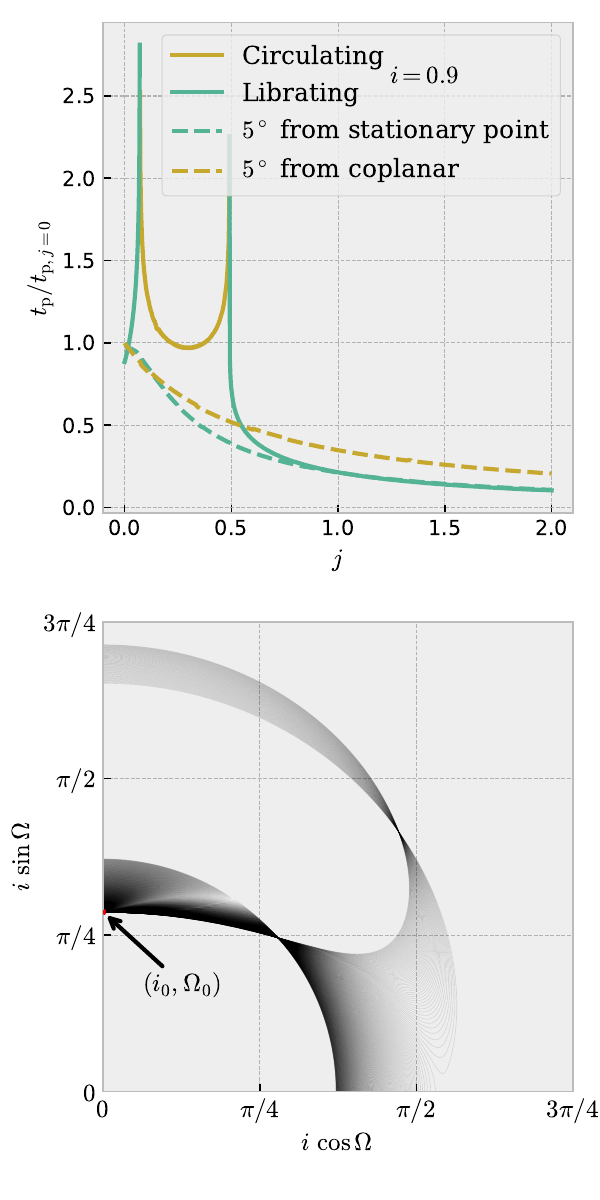}
    \caption{
        Effect of $j$ on the precession timescale as calculated in equation (\ref{eq:tp-rat}). {\bf Top}: Ratio between the numerically computed precession timescale and the analytic solution for a massless ring. To compute the solid lines, the inclination is fixed and only $j$ varies. Twice the orbit switches between \RP{prograde} circulation and libration, and the timescale becomes very long at that boundary. For the dashed lines, $i$ is held $5^\circ$ away from a stationary point. In the librating case, this stationary point is given by equation (15) of \citet{martin2019}. In the circulating case, the stationary point is coplanar. For all simulations the step size is $\Delta j = 10^{-3}$, and we assume $e_\text{b}=0.4$, $f=0.5$, $a_\text{b}/r = 1/5$, and $\Omega_0 = \pi/2$. {\bf Bottom}: $(i,\Omega)$ tracks for all 2000 simulations that compose the solid line in the top panel. The arrow points to the initial conditions for every line (marked by the red point). The divergence of the different tracks is due to the change of $j$ only.
    }
    \script{precession_time.py}
    \label{fig:t_p}
\end{figure}

Figure \ref{fig:t_p} shows that $j$ has two different effects on $t_\text{p}$. The most obvious is that when $j$ changes, the location of the boundary between dynamical regimes also changes. This can cause huge spikes in the timescale ratio as that boundary moves close to the initial conditions $(i_0, \Omega_0)$. \RR{The bottom panel of Figure \ref{fig:t_p} shows that these spikes correspond to transitions in the family of solution. Separating these families (but not shown) are solutions for which the timescales approach infinity.}
The other effect is that the timescale ratio becomes small as $j$ increases. We find that the slope for large j is rather consistent, with $t_\text{p}\propto j^{-1} \text{ to } j^{-1.2}$, with the exponent varying slightly with changes to $f$ or $a_\text{b}/r$. 

Given some set of assumptions, we can calculate the fraction of binary systems for which the precession timescale is too long to reach a coplanar or polar orbit. Planets formed in these disks will undergo secular oscillations for the entire lifetime of their system. This fraction is a strong function of the radius of the ring, and differential precession of an extended disk will cause warps and breaks (see Section \ref{subsec:breaks})

\begin{figure}
    \includegraphics[width=0.5\textwidth]{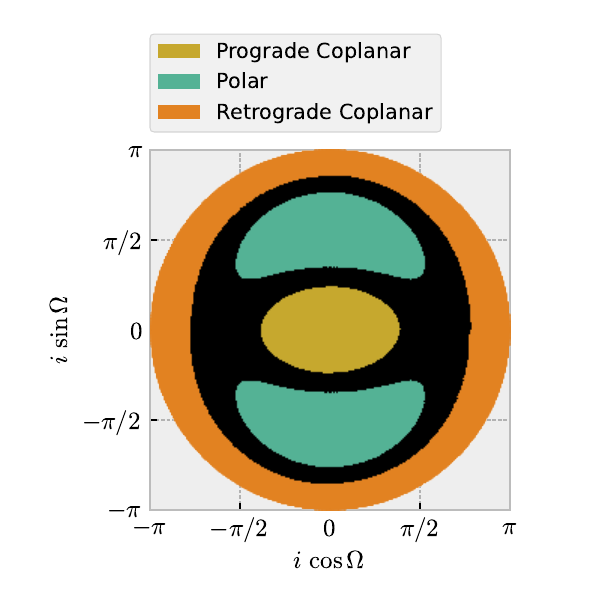}
    \caption{Same as Figure \ref{fig:grid_example}, but with a mask over the region where $t_\text{p} > t_\text{disk}$. Parameters used are an equal-mass binary with $M_\text{b} = M_\odot$, $e_\text{b} = 0.4$, $j=0.1$, $a_\text{b} = 1\text{ AU}$, $r/a_\text{b} = 30$, and $t_\text{disk} = 10^6 \text{ years}$. Note that %
  \input{output/grid_example_tp.txt}\unskip\label{output/grid_example_tp.txt}\unskip%
\% of initial conditions remain misaligned in this setup, but this fraction is near zero when $r/a_\text{b} = 20$ and near 1 when $r/a_\text{b} = 40$. Not shown are versions of this figure where $e_\text{b}$ and $j$ have been varied. We have found that up to 85\% of simulations remain misaligned when $e_\text{b}=10^{-6}$, but if $j$ is subsequently increased to 10 then that becomes 1\%. }
    \script{grid_example_tp.py}
    \label{fig:grid-tp}
\end{figure}

\subsection{Disk breaking}
\label{subsec:breaks}
\citet{rabago2024} provides a framework for discussing extended disks which form discontinuities in $i$ and $\Omega$ due to differential precession and libration. Essentially, the dominant timescale in such a disk is that at the inner edge (recall in equation (\ref{eq:tp-j}) that $t_\text{p} \propto r^{7/2}$).
In a wave-like (low-viscosity) disk, the break occurs at the radius where the inner edge precession timescale is equal to the radial communication timescale. In the diffusive (high-viscosity) disk regime, however, the alignment timescale is very short compared to the precession timescale, and it is possible for the disk to reach coplanar or polar alignment without very much differential precession, suppressing any potential warps. It is still possible, however, for this fast alignment to take longer than a disk lifetime, and we cannot guarantee that all diffusive disks will be able to evolve to coplanar or polar -- only that they are less likely to break.

Consider now a wave-like disk with inner radius $r_\text{in}$ and outer radius $r_\text{out}$. If $r_\text{out}$ is sufficiently large, then a break will occur at a radius $r_\text{break}$ that depends on the specific parameters of the disk \citep[see][eq. (9)]{rabago2024}. The two regions can be treated independently of each other. For the inner disk, the outer edge is sufficiently connected to the inner edge, and the whole region precesses together on a characteristic timescale $t_\text{p}(r_\text{in}) < t_\text{p} \lesssim t_\text{p}(r_\text{break})$. The outer disk may break again if $r_\text{out}$ is very large, but either way it will precess on its own characteristic timescale. We now have two independent rings, each with its own precession timescale. Due to the strong dependence of $t_\text{p}$ on $r$, the inner disk is much more likely to reach alignment in a disk lifetime, and the different alignment timescales means the disks will be misaligned from each other, potentially causing a shadow to be cast on the outer disk \RR{\citep[see][]{bohn2022,benisty2023,su2024,zhang2024,zhu2024}}. \RR{If we choose to measure the polar fraction using the inner ring, then disk breaks will have little effect on $f_\text{polar}$. In practice, however, observed values of $f_\text{polar}$ may depend on whether the technique used to probe the disk configuration is more sensitive to the inner or outer disk.}

\subsection{Self-gravitation}
\label{subsec:toomre}
Just as the disk-breaking framework in Section \ref{subsec:breaks} places an upper limit on the outer radius of an extended disk, we can use the gravitational instability criterion to place a lower limit on that outer radius for constant $j$. In this section we will calculate the maximum $j$ that is allowed for a disk with a given radial extent before it becomes gravitationally unstable. This discussion applies to any disk that one may wish to describe with the dimensionless relative angular momentum $j$, not just the misaligned disks we have focused on so far. The criterion for stability against self-gravitation \citep{toomre1964} can be written
\begin{equation}
    \label{eq:toomre}
    \frac{hM_\text{b}}{\pi r^2 \Sigma} > 1\, ,
\end{equation}
where $h\equiv H/r$ is the aspect ratio of the disk that is assumed to be constant and $\Sigma$ is the surface density. If $\Sigma = \Sigma_\text{in} (r/r_\text{in})^{-p}$ and $p>0$, then the angular momentum of the ring (in the limit $M_\text{r} \ll M_\text{b}$) is
\begin{align}
    J_\text{r} &= \int_{r_\text{in}}^{r_\text{out}} 2\pi r \Sigma  \sqrt{GM_\text{b}r}\,dr \, , \nonumber \\
    & = \frac{2\pi\sqrt{GM_\text{b}}}{5/2 - p} \Sigma_\text{in} r_\text{in}^{5/2} \left( \left( \frac{r_\text{out}}{r_\text{in}} \right)^{5/2 - p}- 1\right)\, \label{eq:Jr}.
\end{align}

First, we rewrite the stability criterion given in equation (\ref{eq:toomre}):
\begin{equation}
    \label{eq:toomre2}
    \frac{h M_\text{b}}{\pi r_\text{in}^p \Sigma_\text{in} r^{2-p}} > 1\, .
\end{equation}

Looking at equation (\ref{eq:Jr}) and equation (3) of \citet{abod2022}, we can write $j$ as
\begin{equation}
    \label{eq:j-full-disk}
    j = \frac{2\pi \Sigma_\text{in} r_\text{in}^{5/2}}{(5/2-p)f(1-f)M_\text{b}\sqrt{a_\text{b}(1-e_\text{b}^2)}}  \left( \left( \frac{r_\text{out}}{r_\text{in}} \right)^{5/2 - p}- 1\right) \, .
\end{equation}

The value of $p$ is typically taken to be less than 2, so the instability onset occurs at the outer edge of the disk. We use equation (\ref{eq:toomre2}) to find a maximum value for $\Sigma_\text{in}$:
\begin{equation}
    \Sigma_\text{max} = \frac{hM_\text{b}}{\pi r_\text{in}^2 (r_\text{out}/r_\text{in})^{2-p}}\, .
\end{equation}

The maximum value of $j$ to be stable against self-gravity is
\begin{multline}
    \label{eq:jmax-shallow}
    j_\text{max} = \frac{2h}{f (1-f) (5/2 - p)\sqrt{1-e_\text{b}^2}}\\ \sqrt{\frac{r_\text{in}}{a_\text{b}}} \left(\frac{r_\text{out}}{r_\text{in}}\right)^{p-2} \left( \left( \frac{r_\text{out}}{r_\text{in}} \right)^{5/2 - p}- 1\right)\, .
\end{multline}

If we consider a $p=1$ disk orbiting an equal mass binary with $e_\text{b}=0.4$, $h=0.01$, and $r_\text{in}/a_\text{b}=2.5$, then for a relatively narrow ring ($r_\text{out}/r_\text{in} = 2$), we calculate $j_\text{max} = 0.08$. If the disk is allowed to be more extended (in this case $r_\text{out}/r_\text{in} = 10$), then we find $j_\text{max} = 0.28$.

The self-gravitation of a disk places an upper limit on what values of $j$ are physical in a given system. As a consequence, a high-$j$ disk is either gravitationally unstable or highly radially extended, although we note that gravitational instability in a warped disc may be suppressed by the heating from the warp propagation \citep{rowther2022}.  A radially extended disk is more likely to break, effectively lowering the value of $j$. This process is highly dependent on the parameters of a particular system, but there may be a general preference for small $j$ by means of gravitational instability and disk breaking.

\RR{
    When a disk does become gravitationally unstable, fragments form and heat up. If the cooling time is greater than a few times the orbital period then fragments collide before they can collapse, leading to a turbulent steady-state with an enhanced effective $\alpha$ viscosity \citep{gammie2001}. Increasing $\alpha$ shortens the evolutionary timescale, so, just like disk breaking, gravitational instability in this regime can result in faster alignment and more disks that reach their polar or coplanar steady-state.
}

\subsection{Effects of companions}
\RR{
    In this work we have only considered disks orbiting isolated binary star systems. However, we can qualitatively extend our results to hierarchical triple star systems. There are two such cases to consider: 1) the outer-companion case where a third star orbits exterior to the circumbinary disk, and 2) the inner-companion case, where the disk orbits a triple-star system. In both the outer \citep[see][]{martin2022,ceppi2023} and inner \citep[see][]{lepp2023,lepp2025} cases, additional conditions are imposed for libration to occur, and $f_\text{polar}$ decreases. The magnitude of these effects is highly dependent on the geometry of the triple-star system.
}

\section{Conclusions}
\label{sec:conclusions}

In this study we attempt to quantify the fraction of polar disks in circumbinary systems using the dynamics of a narrow ring. We do this analytically, numerically in the secular approximation, and numerically using the full binary potential. We find that the three methods produce values of the polar fraction that are identical within their uncertainties.

Our results show that $f_\text{polar}$ is heavily dependent on the underlying parameter distributions that make up a population of systems. In general, we find for low mass disks that $f_\text{polar} \sim e_\text{b}$, and for high mass disks $f_\text{polar} \rightarrow \text{const.}$ as $j\rightarrow \infty$, with the value of the constant being 0.37 if the directions of the disk angular momentum vectors are isotropically distributed.
If we consider a population of low mass disks with a uniform distribution of $e_\text{b}$, then our results are consistent with previous results from \citet{ceppi2024} which found $f_\text{polar}\sim 0.5$. 

Most of the calculations presented in this work assume that the initial disk angular momentum vector has no preference for its direction, and that the distribution is isotropic. However, we also include calculations that unsurprisingly show that $f_\text{polar}$ is reduced when disks tend to be aligned with their host at the time of formation.

Each of the three methods we employ in this paper has its own benefits. With our analytic results, it is possible to compute $f_\text{polar}$ for an arbitrarily large population of systems in less than 1 second. In the same time, our secular numerical method can explore $\sim 50\,000$ combinations of initial conditions for a system, returning relevant timescales and the evolution of orbital parameters for each setup. Our numerical simulations in the full binary potential (using {\sc rebound}) are much slower, \RR{able to determine the dynamical state of a system in $\sim 1\,\text{s}$}, but allow the inclusion of smaller effects such as those from the binary mass fraction and semi-major axis ratios. Each method balances information output with computational cost.

A fourth method, which we do not consider due to its high cost, would be to run full hydrodynamic simulations of the disk. Such simulations would account for all the additional effects discussed in Section \ref{sec:discussion}, but are not suitable for a large parameter space study. In a detailed study of a particular system our results can be used to gain insight into the dynamics before simulating the full disk. For example, equation (\ref{eq:j-full-disk}) can be used to compute $j$ for a power-law disk, and any of the methods described in Section\RR{s} \ref{sec:analytic}-\ref{sec:reb} can then be used to estimate the dynamical state of the system.

We also discuss the limitations of our methods. First, we look at the effect of finite disk lifetime on the ability for a disk to complete its evolution to a stationary state. Disks with long evolutionary timescales remain with orbital parameters near their initial conditions for their entire lifetimes. We also look at limitations imposed by the fact that real disks are not geometrically narrow rings, but have some radial extent. Real disks are likely to break if their radial extent is too large, and will become gravitationally unstable if all of their mass in concentrated in too narrow of a ring. These two effects \RR{shorten the evolutionary timescale and make it more likely that a binary system will host a disk which has reached its stationary configuration, but do not necessarily affect $f_\text{polar}$ as defined in this study}.

\RR{
These results for $f_\text{polar}$ can be tested through population-level observations of circumbinary disk systems. In practice, however, the mutual inclination between the disk and its binary host can be difficult to constrain, and therefore polar disks can be difficult to distinguish from coplanar. Known misaligned disks rely on fortunate geometry to constrain the mutual inclination, such as an eclipse of a nearby binary in the case of V773 Tau B \citep{kenworthy2022}. However, $f_\text{polar}$ could be statistically constrained by looking for polar disks around young eclipsing binaries because the binary-observer inclination is known. Imaging of a face-on disk in such systems would confirm a polar configuration. However, for an edge-on disk it would be difficult to differentiate between a coplanar and polar due to a degeneracy in viewing angle. A carefully done survey of a large population of eclipsing binaries \citep[e.g., those found by the Kepler Mission,][]{kirk2016} would likely constrain the fraction of polar circumbinary disks.}

\RR{
$f_\text{polar}$ could also be inferred from the frequency of polar planets. \citet{zhang2019} show that retrograde apsidal precession of a binary can be caused by a polar circumbinary planet and develop a method to detect such precession in eclipsing binaries via eclipse timing variations. \citet{baycroft2025} successfully detect retrograde apsidal precession using radial velocity measurements, and conclude that a polar planet is the only suitable explanation. Both of these methods could be applied to the population of known eclipsing binaries. Compared to the imaging method suggested previously, planet-focused methods would be most sensitive to small separations from the binary, meaning the alignment timescale for the progenitor disk would be much shorter than for those can can be resolved via imaging. Together, however, these two methods (disk imaging, apsidal precession inference) could provide robust constraints of $f_\text{polar}$ across many regimes of host parameters and disk separation.
}

\vspace{1cm} 

We acknowledge support from NASA through grants 80NSSC21K0395 and 80NSSC19K0443. This manuscript was prepared using the open-science software \href{https://show-your.work/en/latest/intro/}{\showyourwork} \citep{luger2021}, making the article completely
reproducible. The source code to compile this document and create the figures is available on GitHub\footnote{\ghurl}.
Simulations in this paper made use of the {\sc rebound} N-body code \citep{rebound}.
The simulations were integrated using IAS15, a 15th order Gauss-Radau integrator \citep{reboundias15}. 

\pagebreak
\appendix

\section{Computing $\Gamma_2$}
\RR{
Equation (3.9) from \citet{farago2010} includes a term $\sqrt{1 + m_\text{r}/M_\text{b}}$ where $m_\text{r}$ is the mass of the ring and $M_\text{b}$ is the mass of the binary host. To find the value of this expression in our own variables is non-trivial. Here we define
\begin{equation}
    \Gamma_2 = \sqrt{1 + \frac{m_\text{r}}{M_\text{b}}}\, .
\end{equation}
}

\RR{
We can write
\begin{equation}
    \label{eq:gamma2-eq}
    \Gamma_2^3 - \Gamma_2 - j\,f\,(1-f)\,\sqrt{\frac{a_\text{b}}{r}\,(1-e_\text{b}^2)} = 0\, ,
\end{equation}
and solve for $\Gamma_2$ using the Newton-Raphson method.
}

\bibliography{cbdisk,syw,reb}

\end{document}